\documentclass[preprint,prb]{revtex4}

\usepackage{graphicx}
\usepackage{dcolumn}
\usepackage{bm}
\usepackage{natbib}
\usepackage[english]{babel}
\usepackage{amsmath}
\usepackage{float}

\begin{document}

\title{\boldmath Fermi liquid-like behaviour of cuprates in the pseudogap phase simulated via $T$-dependent electron-boson spectral density \unboldmath}

\author{Hwiwoo Park} \author{Jungseek Hwang}
\email{jungseek@skku.edu}
\affiliation{Department of Physics, Sungkyunkwan University, Suwon, Gyeonggi-do 16419, Republic of Korea}

\date{\today}

\begin{abstract}

We investigated the temperature- and frequency-dependent optical scattering rates in the pseudogap phase of cuprates using model pseudogap and electron-boson spectral density (EBSD) functions. We obtained the scattering rates at various temperatures below and above a given pseudogap temperature using a generalized Allen’s (or Sharapov’s) formula, which has been used to analyse the measured optical spectra of correlated electron systems with a non-constant density of states at finite temperatures. The pseudogap and EBSD functions should be temperature dependent to simulate the Fermi liquid-like behaviour of underdoped cuprate systems observed in optical studies. Therefore, the observed Fermi liquid-like behaviour can be understood by considering the combined contribution from the $T$-dependent EBSD function and the $T$-dependent pseudogap. We also obtained the optical conductivity spectra from the optical scattering rates and analyzed them to investigate intriguing electronic properties. We expect that our results will aid in understanding the Fermi liquid-like optical response in the pseudogap phase and in revealing the microscopic pairing mechanism for superconductivity in cuprates.
\\

\noindent *Correspondence to [email: jungseek@skku.edu].

\end{abstract}


\maketitle

\section{Introduction}

Copper oxide superconductors (or cuprates) have been studied intensively and most thoroughly since their discovery \cite{bednorz:1986}. The temperature-doping phase diagram has become increasingly complicated \cite{keimer:2015}. Each characteristic phase region in the phase diagram is still not fully understood. The pseudogap \cite{timusk:1999} is the most studied phase among the phase regions. However, the origin of the pseudogap is not well understood. The origin of the pseudogap is crucial for understanding the superconductivity in cuprates. A study \cite{mirzaei:2013} claimed to have found spectroscopic evidence for Fermi liquid-like behaviour in the temperature- and frequency-dependent relaxation rate of heavily underdoped single crystals of HgBa$_2$CuO$_{4+\delta}$ (Hg1201) with $T_c$ =67 K. This Fermi liquid-like behaviour occurs at low temperatures below the onset temperature of the pseudogap phase ($T^{*}$) and at low frequencies below the pseudogap energy ($\Delta_{pg}$). These findings can provide important information to better understand the metallic state and high-temperature superconductivity in cuprates. There are some recent intriguing optical studies on Fermi-liquid-like behavior and the strange (or bad) metal states of cuprates \cite{heumen:2022,kumar:2022}. The authors showed that the optical conductivity in the low-energy region consisted of two components, which are the Drude component and the mid-infrared absorption one. They analyzed the measured optical spectra to reveal temperature- and doping-dependencies of these components over a wide doping range, from underdoped to overdoped.

In our study, we mainly focus on simulating the measured optical spectra of a undedoped cuprate starting from a model electron-boson spectral density (EBSD) function using the generalized Allen's (or Sharapov's) formula \cite{allen:1971,sharapov:2005} and the extended Drude model \cite{gotze:1972,allen:1977,puchkov:1996,hwang:2004}. We used a well-established phenomenological approach \cite{allen:1971,shulga:1991,sharapov:2005,schachinger:2006} to simulate the Fermi liquid-like behaviour of cuprates in the pseudogap phase. This approach has been used to extract the electron-boson spectral density functions (or information on correlations) from measured optical spectra of cuprates \cite{allen:1971,shulga:1991,sharapov:2005,schachinger:2006,hwang:2006,hwang:2007,schachinger:2008,hwang:2011}. We started with a model electron-boson spectral density (EBSD) function \cite{hwang:2011} and a pseudogap model \cite{hwang:2008} for a given temperature, and obtained a corresponding optical scattering rate (or the imaginary part of the optical (or two-particle) self-energy) using a generalized Allen's (or Sharapov's) formula \cite{sharapov:2005}, which can be used for analyzing a measured spectrum of a correlated electron system with a non-constant density of states, such as the pseudogap in cuprates \cite{hwang:2006}, and at a finite temperature. We then used the Kramers-Kronig relation between the imaginary and real parts of the optical energy \cite{hwang:2015a} to obtain the corresponding real part of the optical self-energy. Furthermore, we obtained the optical conductivity from the complex optical self-energy using the extended Drude model formalism \cite{gotze:1972,allen:1977,puchkov:1996,hwang:2004}. We used two different sets of the EBSD functions (one is temperature dependent, and the other temperature independent) and found that the experimentally observed optical scattering rate could be simulated with the temperature-dependent EBSD function. The results showed that the temperature dependencies of the model EBSD function and model pseudogap are crucial in describing the observed optical response of cuprates in the pseudogap phase. Therefore, this can be a piece of important evidence that the EBSD function intrinsically contains temperature dependency, which may be critical in understanding superconductivity in cuprates. Our results will help us to understand the Fermi liquid-like optical response in the pseudogap phase and to find out the elusive microscopic pairing mechanism for superconductivity in cuprates.

\section{Theoretical formalism}

Information regarding the correlations between charge carriers in a correlated electron system is encoded in the measured optical spectrum. In more detail, because the occupied and unoccupied electron states are involved in the optical spectrum, the band renormalization caused by the correlations will naturally appear in the measured optical spectrum, which contains intraband and interband transitions from filled states to empty states and is governed by the joint density of states and the dipole selection rule. The encoded correlation information can be extracted from the measured optical spectrum using a well-established approach \cite{allen:1971,shulga:1991,gotze:1972,allen:1977,puchkov:1996,hwang:2004,sharapov:2005,schachinger:2006}. The established approach consists of a couple of formalisms, such as the extended Drude formalism and the generalized Allen formalism. The correlated carriers can be described by the extended Drude formalism, whereas the free electrons (or Fermi gas) can be described by the Drude formalism. In the extended Drude formalism \cite{gotze:1972,allen:1977,puchkov:1996,hwang:2004}, the complex optical conductivity ($\tilde{\sigma}(\omega)$) can be written as follows:
\begin{equation}\label{eq1}
  \tilde{\sigma}(\omega,T) = i\frac{\Omega_p^2}{4\pi}\frac{1}{\omega+[-2\tilde{\Sigma}^{op}(\omega,T)]}
\end{equation}
where $\Omega_p$ is the plasma frequency of the itinerant electrons. Here the impurity elastic scattering rate in the Drude model is simply replaced with a frequency-dependent quantity to describe the energy transfer between the charge carriers caused by the correlations and the frequency-dependent quantity should be a complex function to maintain the causality. The complex quantity ($\tilde{\Sigma}^{op}(\omega,T)$) is called the optical self-energy and carries the information of correlations between charge carriers. The real and imaginary parts of the optical self-energy form a Kramers-Kronig pair. In principle, the correlations may result in band renormalizations, and the band renormalizations can be encoded in the measured optical spectrum. Therefore, the information on correlations can be encoded in the complex optical self-energy, which is the deviation from the Drude band or, more generally, the bare band. The bare band can be obtained using the local density approximation calculations. The optical self-energy is associated with a two-particle (or optical) process because both the filled and empty states (or electrons and holes) are involved in the optical (or absorption) process. Furthermore, the optical self-energy is closely related to the quasiparticle self-energy ($\tilde{\Sigma}^{qp}(\omega)$), which is associated with a single particle process because either only the filled states (or electrons) or the empty states (or holes) are involved in the process. Therefore, the optical self-energy is more complicated and can be related to the quasiparticle self-energy as, $\tilde{\Sigma}^{qp}(\omega) = \frac{d[\omega \tilde{\Sigma}^{op}(\omega)]}{d\omega}$ \cite{hwang:2007b}. In principle, the quasiparticle self-energy can be measured using angle-resolved photoemission spectroscopy (ARPES), whereas the optical self-energy can be measured using infrared/optical spectroscopy. We note that using ARPES, the quasiparticle self-energy of only the filled states (or electrons) can be obtained because of the experimental limitation; only the photoelectrons from the filled states can be measured with the technique.

A force-mediated boson exchanging model can be used to describe the correlations. In this model, the correlations can be described in terms of the EBSD function, $I^2B(\Omega)$, where $I$ is the coupling constant between an electron and a boson and $B(\Omega)$ is the boson spectrum. In this case, the optical self-energy can be expressed in terms of the EBSD function. Allen originally developed an expression \cite{allen:1971}, which can be used for the case of the normal state at $T$ = 0 and the $s$-wave superconducting state. The Allen's expression has been generalized by Shulga {\it et al.} \cite{shulga:1991} for use in the case of a finite temperature, and further generalization has been achieved by Sharapov {\it et al.} for use in the case of a finite temperature and a non-constant density of states, such as the pseudogap phase in underdoped cuprates \cite{sharapov:2005}. This generalized Allen formalism \cite{sharapov:2005} can be written as follows:
\begin{eqnarray}\label{eq2}
1/\tau^{op}(\omega,T) &=& \frac{\pi}{\omega} \int_0^{\infty}d\Omega \: I^2B(\Omega,T)\int_{-\infty}^{+\infty} dx[N(x-\Omega,T) \nonumber \\
 &+& N(-x+\Omega,T)] [n_B(\Omega,T)+f(\Omega-x,T)] \nonumber \\
 && [f(x-\omega,T)-f(x+\omega,T)],
\end{eqnarray}
where $n_B(\omega,T)$ and $f(\omega,T)$ are the Bose-Einstein and Fermi-Dirac distribution functions, respectively, and $N(x,T)$ is the normalized density of states, which is used to describe the pseudogap. The pseudogap (PG) can be modelled \cite{hwang:2008b,hwang:2012} as follows:
\begin{eqnarray}\label{eq3}
  N(x,T)&=& N_0(T)+[1-N_0(T)] \Big{(} \frac{x}{\Delta_{pg}} \Big{)}^2 \:\:\mbox{for}\:\: |x|\leq\Delta_{pg}, \nonumber  \\
  &=& 1+\frac{2[1-N_0(T)]}{3} \:\:\mbox{for}\:\: |x|\in(\Delta_{pg}, 2\Delta_{pg}), \nonumber \\
  &=& 1 \:\:\mbox{for}\:\: |x|\geq 2\Delta_{pg},
\end{eqnarray}
where $N_0(T)$ is the normalized density of states at the Fermi level and $\Delta_{pg}$ is the size of the pseudogap. This pseudogap model has been used for analyzing the optical spectra of underdoped cuprate systems \cite{hwang:2008,hwang:2008b,hwang:2011,hwang:2012,hwang:2013}. For this pseudogap model, the density of states loss in the pseudogap is fully recovered just above the pseudogap within $2\Delta_{pg}$ \cite{hwang:2008,hwang:2008b}. For Fermi surface arcs, the normalized density of states at the Fermi level is proportional to the temperature \cite{hwang:2008b}, i.e. $N_0(T) = T/T^*$, where $T^*$ is the pseudogap temperature. Because the measured optical spectrum is $k$-space averaged, a $k$-space averaged PG model is used, and the Fermi arc model is used for the $T$-dependent depth ($1 - N_0$) of the PG, which is closely associated with the $d$-wave-like PG and may also be related to the $d$-wave SC gap.

The EBSD functions, $I^2B(\Omega,T)$, of cuprates have been obtained using various spectroscopic experimental techniques, such as tunneling spectroscopy, ARPES, optical spectroscopy, Raman spectroscopy, and inelastic neutron scattering \cite{carbotte:2011}. The obtained EBSD functions showed a unified temperature and doping-dependent evolution. A broad spectrum at high temperatures transforms into a peak between 30 and 60 meV and a featureless high-energy background through a spectral weight redistribution as the temperature decreases. The overall spectral weight shifts to a higher energy region and the spectral weight redistribution weaken as the doping increases.\cite{hwang:2018,hwang:2021}. In this study, we used the temperature-dependent model $I^2B(\Omega,T)$. The model $I^2B(\Omega,T)$ consists of two (sharp and broad) components. For the broad component, we used an MMP mode\cite{millis:1990}, which was introduced by Millis, Monien, and Pines (MMP) and was used to describe the antiferromagnetic fluctuations. For the sharp component, we used a sharp mode, which has been previously used to analyze the measured optical spectra of underdoped cuprates \cite{hwang:2011}. The model $I^2B(\Omega)$ can be written as follows:
\begin{eqnarray}\label{eq4}
I^2B(\Omega, T) &=& \frac{2 A_b(T)}{\ln{\Big{[}}1+(\frac{\omega_c}{\omega_b(T)})^2 \Big{]}}\frac{\Omega}{\Omega^2 +\omega_b(T)^2} \nonumber \\ &+& \frac{2 A_s(T) \omega_s(T)^2}{\tan^{-1}\Big{[}(\frac{\omega_c}{\omega_s(T)})^2\Big{]}}\frac{\Omega}{\Omega^4+\omega_s(T)^4}
\end{eqnarray}
where the first term is the broad MMP mode and the second one is the sharp mode. $A_b$ and $\omega_b$ are the area under and peak frequency (or energy) of the broad MMP mode, respectively. $A_s$ and $\omega_s/\sqrt[4]{3}$ are the area under and peak frequency of the sharp mode, respectively, and $\omega_c$ is the cutoff frequency. The area under each mode and the peak frequency of each mode depend upon temperature. Note that $I^2B(\Omega)$ is exactly zero at zero energy in this model.

When we obtain the optical scattering rate, we can go further and get the complex optical conductivity \cite{hwang:2015a}. Because the optical scattering rate is identical to -2 times the imaginary part of the optical self-energy, i.e. $1/\tau^{op}(\omega) \equiv -2\Sigma_2^{op}(\omega)$ and the real and imaginary parts of the optical self-energy form a Kramers-Kronig pair, we can get the real part of the optical self-energy from the imaginary part using the Kramers-Kronig relation between them, i.e. $-2\Sigma^{op}_1(\omega) = -\frac{2}{\pi} P\int^{\infty}_{0}\{\Omega[-2\Sigma^{op}_2(\Omega)]\}/(\Omega^2-\omega^2)d\Omega$ \cite{hwang:2015a}, where $P$ is the principle part of the improper integral. The Kramers-Kronig relation is a Cauchy principal value integral. And then, we obtain the complex optical conductivity from the complex optical self-energy using the extended Drude model, Eq. (1).

\section{Results and discussion}

\begin{figure}[!htbp]
  \vspace*{-0.3 cm}%
 \centerline{\includegraphics[width=7.0 in]{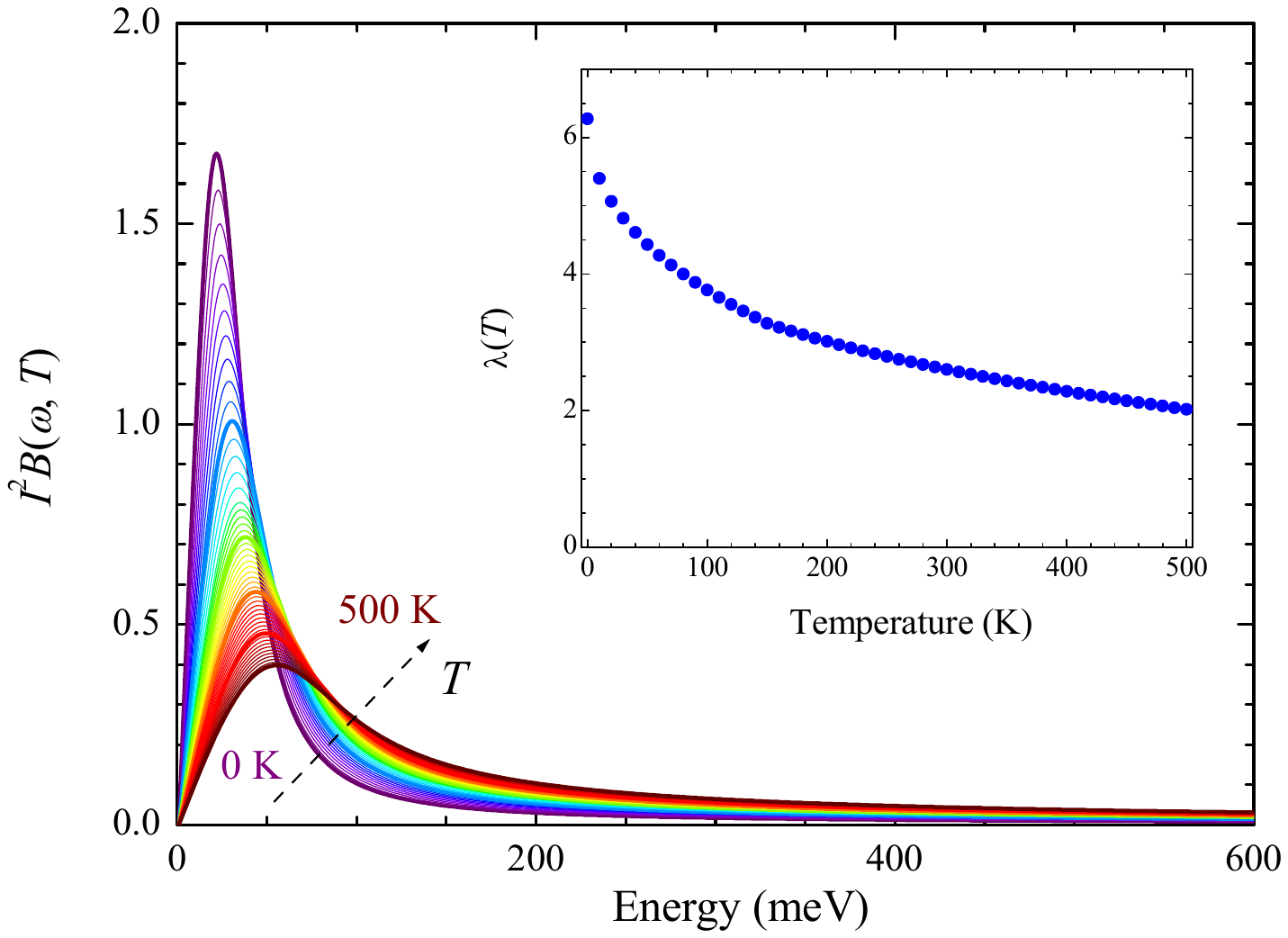}}%
  \vspace*{-0.5 cm}%
\caption{(Color online) Electron-boson spectral density (EBSD) functions at temperatures ranging from 0 to 500 K with a 10 K increment. The EBSD function consists of two components. Note that the thick purple is 0 K, the thick blue curve is 100 K, the thick green curve is 200 K, the thick orange curve is 300 K, the thick red curve is 400 K, and the thick dark brown curve is 500 K. In the inset, the temperature-dependent coupling constant, $\lambda(T)$, is shown.}
\label{fig1}
\end{figure}

Fig. \ref{fig1} shows our model EBSD functions ($I^2B(\Omega)$) at various temperatures ranging from 0 to 500 K with a 10 K increment. The EBSD function consists of two components, namely, sharp and broad modes (Eq. (4)). Note that the sharp mode linearly shifts to a higher energy and its intensity (or area) linearly decreases as temperature decreases, whereas the broad mode linearly shifts to a higher energy and its intensity linearly increases as temperature decreases. As the temperature is reduced, the total area of the two modes gradually decreases. The detailed temperature-dependent changes of the all parameters in the $T$-dependent EBSD function are shown in Fig. S1 in the Supplementary Material. As a result, the EBSD function ($I^2B(\omega, T)$) is $T$-dependent and extends over a very broad spectral range up to 625 meV, whereas the electron-phonon spectral density (EPSD) functions of the conventional phonon-mediated superconductors are $T$-independent and typically extend below 30 meV \cite{carbotte:1990}. It is worth noting that we obtained the $T$-dependent EBSD function by referring to previously published $I^2B(\omega)$ of underdoped cuprates \cite{hwang:2008b,hwang:2011,hwang:2021}. We obtain the coupling constant ($\lambda(T)$) from the model $I^2B(\Omega, T)$, which is defined by $\lambda(T) \equiv 2\int_0^{\omega_c} [I^2B(\Omega,T)/\Omega] d\Omega$, where $\omega_c$ is the cutoff frequency, 625 meV in this study. We show the coupling constant as a function of temperature in the inset of Fig. \ref{fig1}. The coupling constant monotonically increases as the temperature decreases, which is a similar trend observed in underdoped cuprates \cite{hwang:2011,hwang:2016a,hwang:2021}. $\lambda (T)$ is $\sim$ 3-6 for $T \leq$ 200 K, which is much larger than that ($\sim$1.55) of a strong-coupling superconductor, Pb \cite{carbotte:1990}. Because the EBSD function is $T$-dependent, the resulting $\lambda$ is $T$-dependent, whereas that of the conventional phonon-mediated superconductors is $T$-independent. The $T$-dependence and the high values of $\lambda$ are very similar to those of underdoped YBa$_2$Cu$_3$O$_{6+x}$ or Bi$_2$Sr$_2$CaCu$_2$O$_{8+\delta}$ \cite{hwang:2021}. These differences between the EBSD function and the EPSD function of conventional superconductors may indicate that the EBSD function contains a boson that is not a phonon. For cuprates, the force-mediated boson could be antiferromagnetic spin fluctuations because the EBSD function of cuprates exhibits intriguing $T$- and doping-dependencies, which agree with the phase diagram of cuprates; as both temperature and doping decrease, the coupling constant ($\lambda$) significantly increases \cite{hwang:2021}.

\begin{figure}[!htbp]
  \vspace*{-0.3 cm}%
 \centerline{\includegraphics[width=7.0 in]{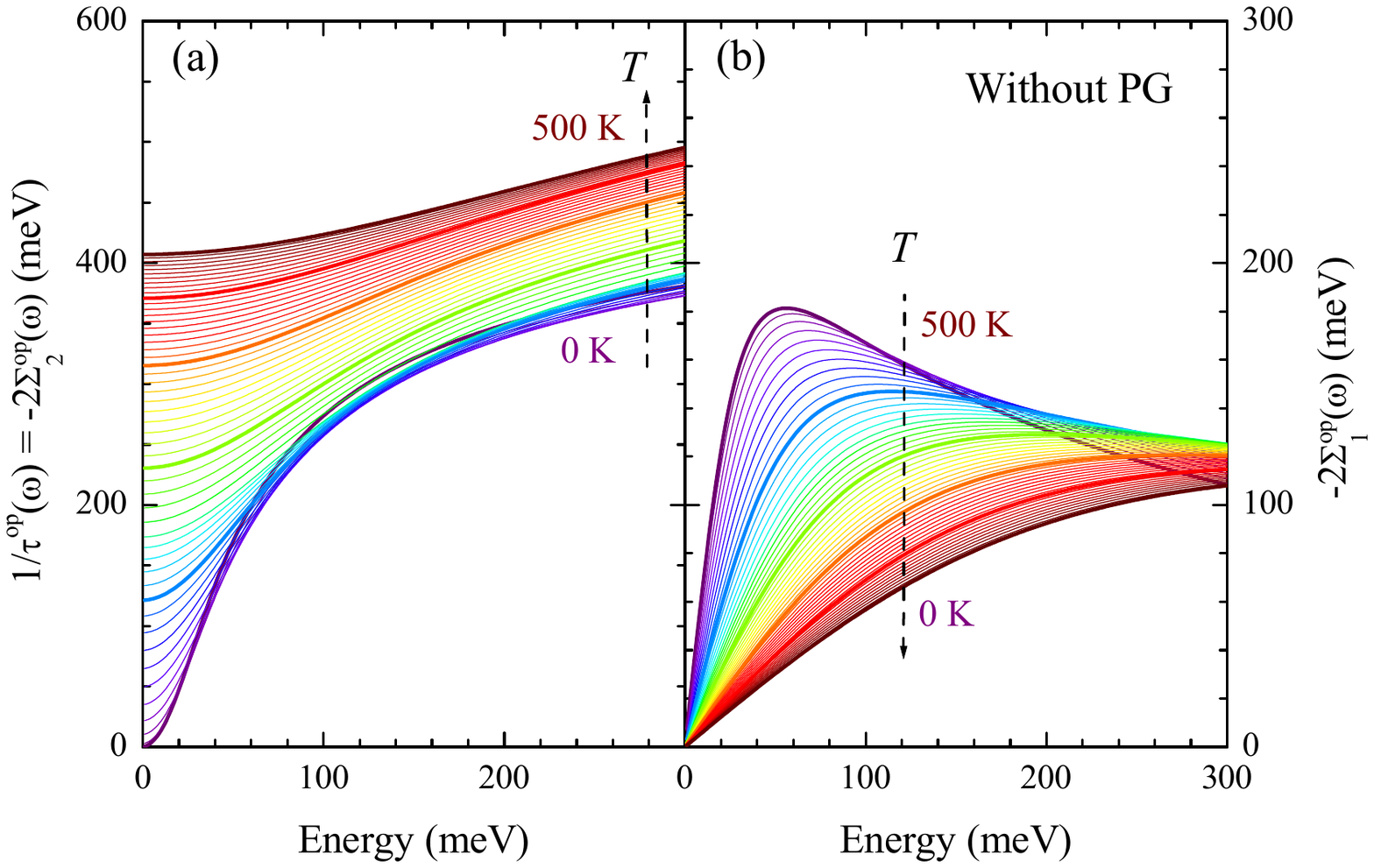}}%
  \vspace*{-0.5 cm}%
\caption{(Color online) Complex optical self-energy, without including the pseudogap. (a) Optical scattering rates (imaginary parts of the optical self-energy) and (b) the real parts of the optical self-energy at temperatures ranging from 0 to 500 K with a 10 K increment. Note that the thick purple is 0 K, the thick blue curve is 100 K, the thick green curve is 200 K, the thick orange curve is 300 K, the thick red curve is 400 K, and the thick dark brown curve is 500 K. Note that the real part was obtained from the imaginary part using the Kramers-Kronig relation between them (see the text for a more detailed description).}
\label{fig2}
\end{figure}

Fig. \ref{fig2}(a) shows the optical scattering rates (or the imaginary parts of the optical self-energy) at various temperatures ranging from 0 to 500 K obtained from the temperature-dependent model $I^2B(\omega,T)$ shown in Fig. \ref{eq1} using Eq. (2) and Eq. (3) with $N_0(T) = 1.0$, i.e. without including the pseudogap. At each temperature, we observe a rapid increase near the peak position of $I^2B(\Omega)$ resulting in a step-like feature. The step-like feature weakens with increasing temperature, which is consistent with the temperature-dependent intensity of the sharp mode (Fig. \ref{fig1}). In most regions, the scattering rate increases monotonically increases as the temperature increases. Note that when the peak of $I^2B(\Omega)$ is very sharp, the order of the scattering rate can be reversed, as seen at low temperatures above $\sim$300 meV. Fig. \ref{fig2}(b) shows the corresponding real parts of the optical self-energy at various temperatures which were obtained from the imaginary parts using the Kramers-Kronig relation between the real and imaginary parts. A systematic temperature-dependent trend is shown. A broad peak is observed at the energy where the maximum slope in the scattering rate is located. The peak is shifted to higher energy as the temperature increases. However, the shape of the peak is different from that of the peak in the measured real part of the optical self-energy of underdoped cuprate systems \cite{hwang:2006,hwang:2008,mirzaei:2013}.

\begin{figure}[!htbp]
  \vspace*{-0.3 cm}%
 \centerline{\includegraphics[width=7.0 in]{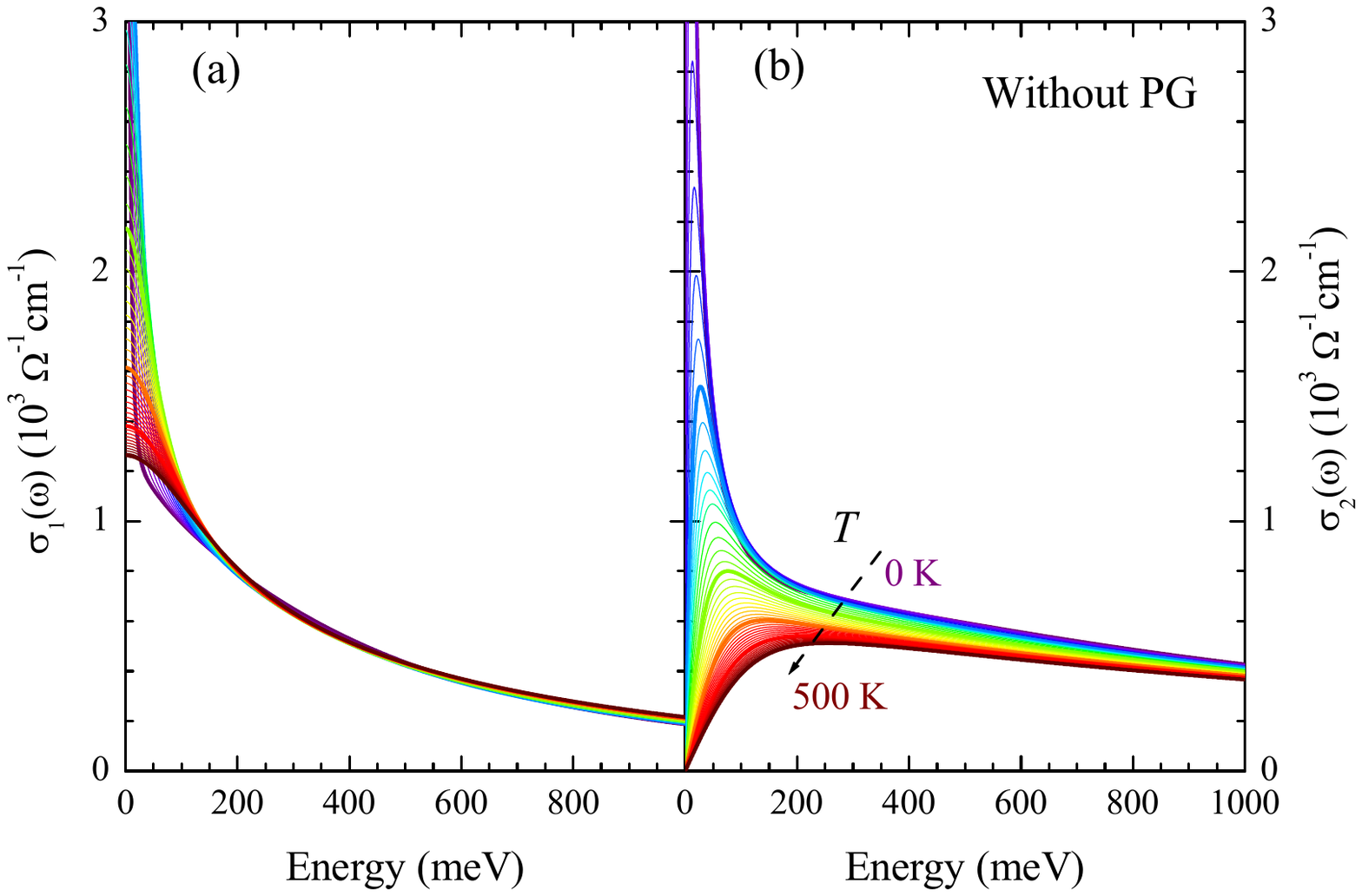}}%
  \vspace*{-0.5 cm}%
\caption{(Color online) Complex optical conductivity, without including the pseudogap. (a) Real and (b) imaginary parts of the optical conductivity at various temperatures ranging from 0 to 500 K with a 10 K increment. The optical conductivity was obtained from the optical self-energy using the extended Drude model with the plasma frequency of 2.0 eV and $1/\tau_{\mathrm{imp}}$ of 15 meV. Here, the impurity scattering rate is independent of both frequency and temperature.}
\label{fig2a}
\end{figure}

\begin{figure}[!htbp]
  \vspace*{-0.3 cm}%
 \centerline{\includegraphics[width=7.0 in]{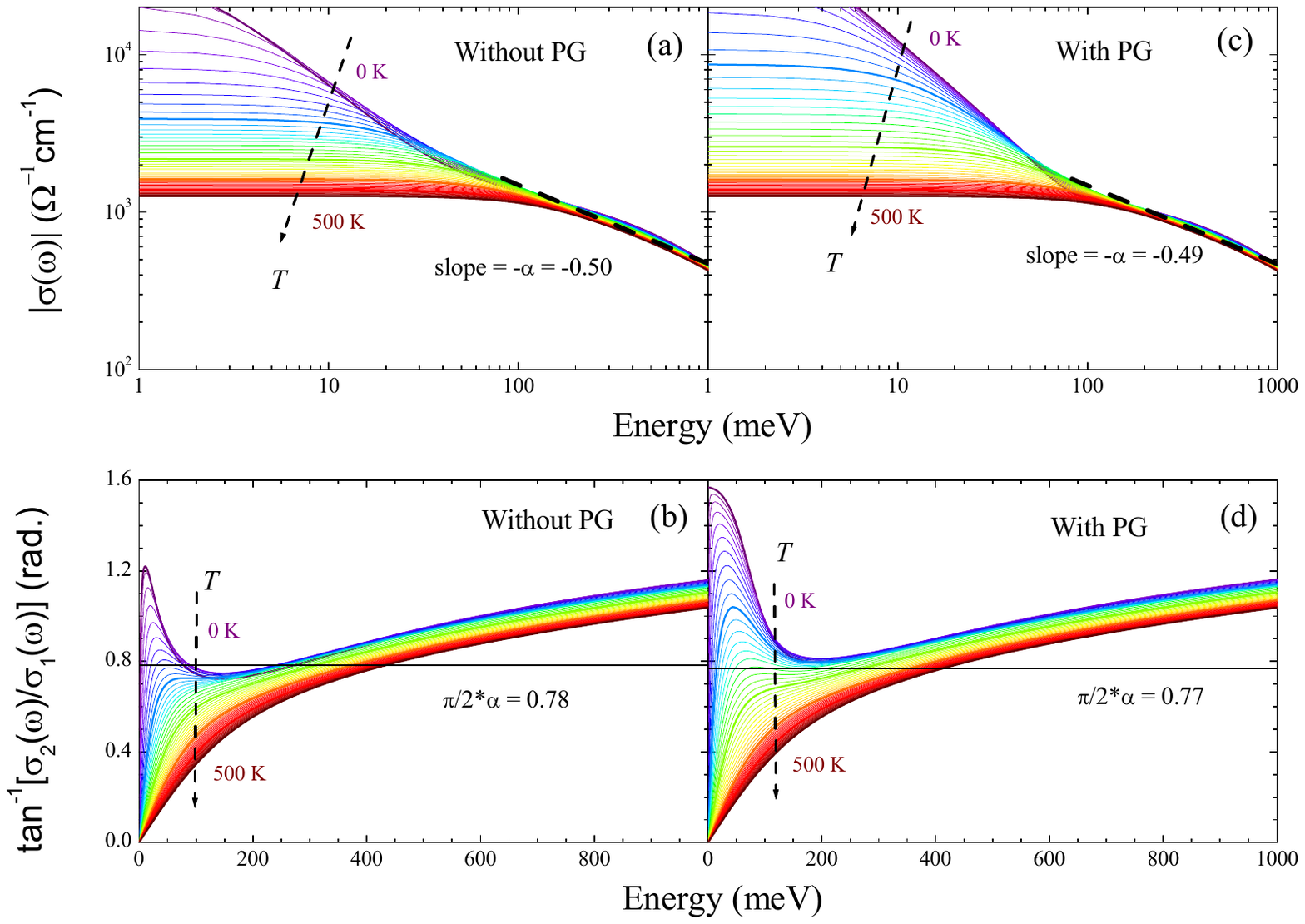}}%
  \vspace*{-0.5 cm}%
\caption{(Color online) (a) The amplitude ($|\sigma(\omega)|$) and (b) the phase angle ($\tan^{-1}[[\sigma_2(\omega)/\sigma_1(\omega)]$) of the optical conductivity for the case without including pseudogap (PG). The slope ($\alpha$) of the linear line is -0.50. To get the slope, we took a linear fit to the data between 0.1 and 1.0 eV at 200 K. The thick dashed line is the liner fit. The solid horizontal line in (b) is $\alpha(\pi/2)$. The results seem to be consistent with those of an underdoped cuprate in a published paper \cite{mirzaei:2013}. (c) The amplitude and (d) the phase angle of the optical conductivity for the case of including PG. They show significant differences in the low-energy region when compared to those when the PG is not included; they show much enhanced conductivity in the low-energy region due to the PG.}
\label{fig2b}
\end{figure}

Figs. \ref{fig2a}(a) and (b) show the real and imaginary parts of the optical conductivity, respectively, which were obtained from the real and imaginary parts of the optical self-energy and the extended Drude formula, Eq. (1). To obtain the optical conductivity, the plasma frequency ($\Omega_p$) was set at 2.0 eV, and the impurity scattering rate ($1/\tau_{\mathrm{imp}}$) of 15 meV was included. The temperature-dependent real part of the optical conductivity is significantly different from that of the measured one in the paper by Mirzaei {\it et al.} \cite{mirzaei:2013}. We also obtained the amplitude ($|\sigma(\omega)| \equiv \sqrt{[\sigma_1(\omega)]^2+[\sigma_2(\omega)]^2}$) and phase angle ($\tan^{-1}[\sigma_2(\omega)/\sigma_1(\omega)]$) of the complex optical conductivity \cite{marel:2003,hwang:2007a,heumen:2022} and showed them in Figs. \ref{fig2b}(a) and (b). Interestingly, the results seem to be consistent with those of underdoped cuprate in a published paper \cite{heumen:2022}, even though the conductivity is much different from that of a measured underdoped cuprate \cite{hwang:2006,mirzaei:2013}, which may indicate that the pseudogap does not affect the conformal tail (or the mid-infrared absorption). In Figs. \ref{fig2b}(c) and (d), we also show the amplitude and phase angle of the optical conductivity when the pseudogap is included for comparison.  The two sets of data are significantly different in the low-energy region due to the PG. However, in the high-energy region, they are almost identical, indicating that the PG influences are confined to the low-energy region.

\begin{figure}[!htbp]
  \vspace*{-0.3 cm}%
 \centerline{\includegraphics[width=7.0 in]{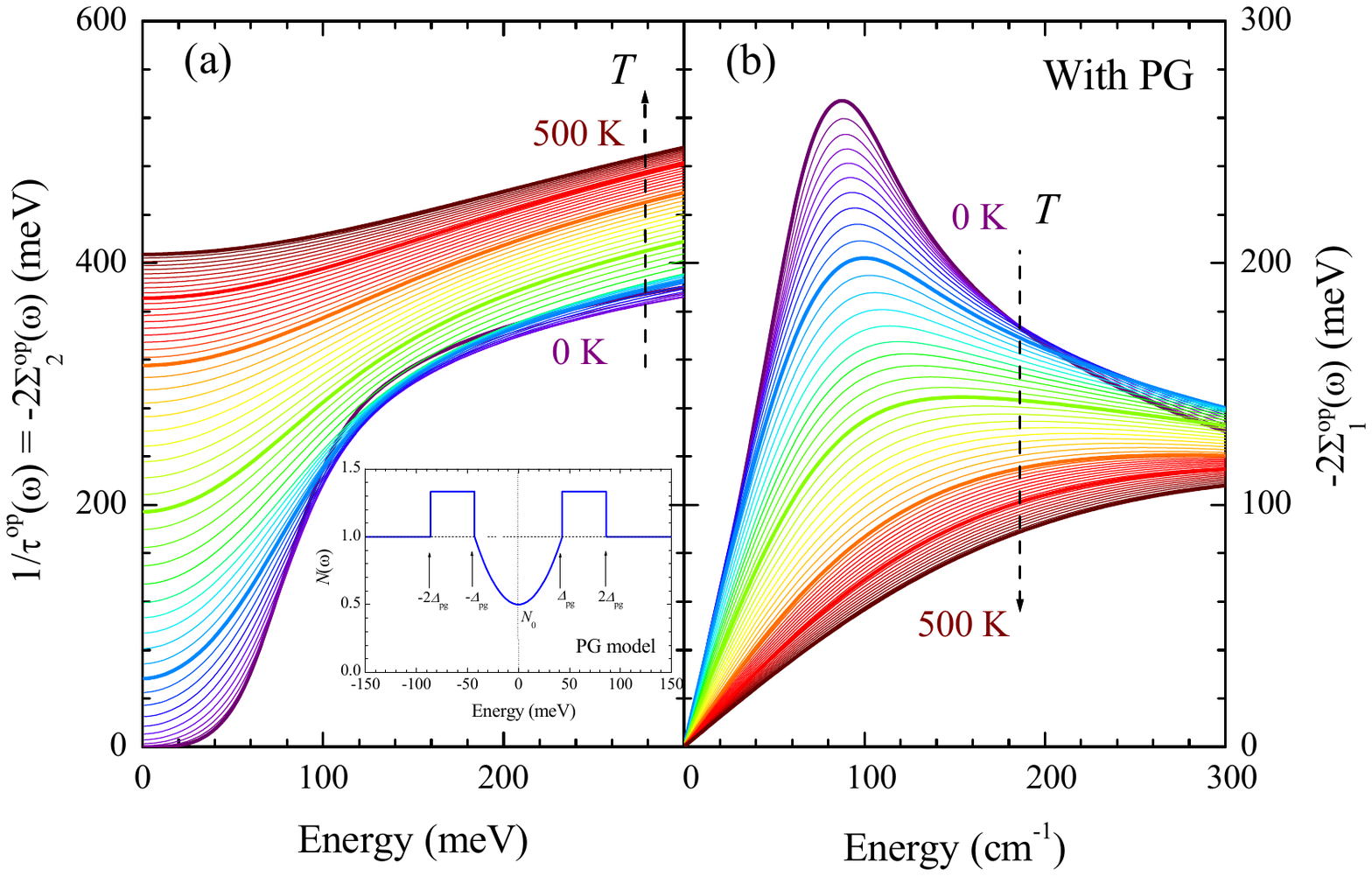}}%
  \vspace*{-0.5 cm}%
\caption{(Color online) Complex optical self-energy, including the pesudogap. (a) Optical scattering rates (or imaginary parts of the optical self-energy) (b) real parts of the optical self-energy at various temperatures ranging from 0 to 500 K with a 10 K increment. In the inset, the model pseudogap is shown for a case of $N_0 =$ 0.5. The real part was obtained from the imaginary part using the Kramers-Kronig relation between them.}
\label{fig3}
\end{figure}
\newpage

Figs. \ref{fig3}(a) and (b), respectively, shows the real and imaginary parts of the optical self-energy at various temperatures ranging from 0 to 500 K obtained from the temperature-dependent model $I^2B(\Omega,T)$ using Eq. (2) and Eq. (3) with $N_0(T) \leq 1.0$, i.e. with including the pseudogap for $T \leq T^*$, where $T^*$ is the pseudogap temperature (in our case $T^* =$ 300 K). To include the temperature-dependent pseudogap, we used the Fermi arc model \cite{hwang:2008b}, i.e. $N_{0}(T) = T/T^*$ for $T \leq T^*$ in Eq. (3) and $N_0(T) = 1.0$ for $T > T^*$. The model pseudogap is shown in the inset of Fig. \ref{fig3}(a) for $N_0 = 0.5$ and $\Delta_{pg} = 43$ meV. Comparing Fig. \ref{fig3}(a) with Fig. \ref{fig2}(a), the step-like feature is more pronounced and shifts to a higher energy because the pseudogap suppresses the density of states near the Fermi level. The reduction of the density of states near the Fermi level results in the reduction of scatterers. Generally, as the pseudogap deepens, the shift of the step-like feature grows larger \cite{hwang:2011}. The step-like feature results from a combined effect of the pseudogap and the sharp peak in $I^2B(\Omega)$\cite{hwang:2012}. Above $T^*$, two sets of the scattering rates shown in Figs. \ref{fig2} and \ref{fig3} are identical. Comparing Fig. \ref{fig3}(b) with Fig. \ref{fig2}(b), the broad peak is shifted to a higher energy and is sharper and better defined than that in Fig. \ref{fig2}(b) due to the pseudogap.

\begin{figure}[!htbp]
  \vspace*{-0.3 cm}%
 \centerline{\includegraphics[width=7.0 in]{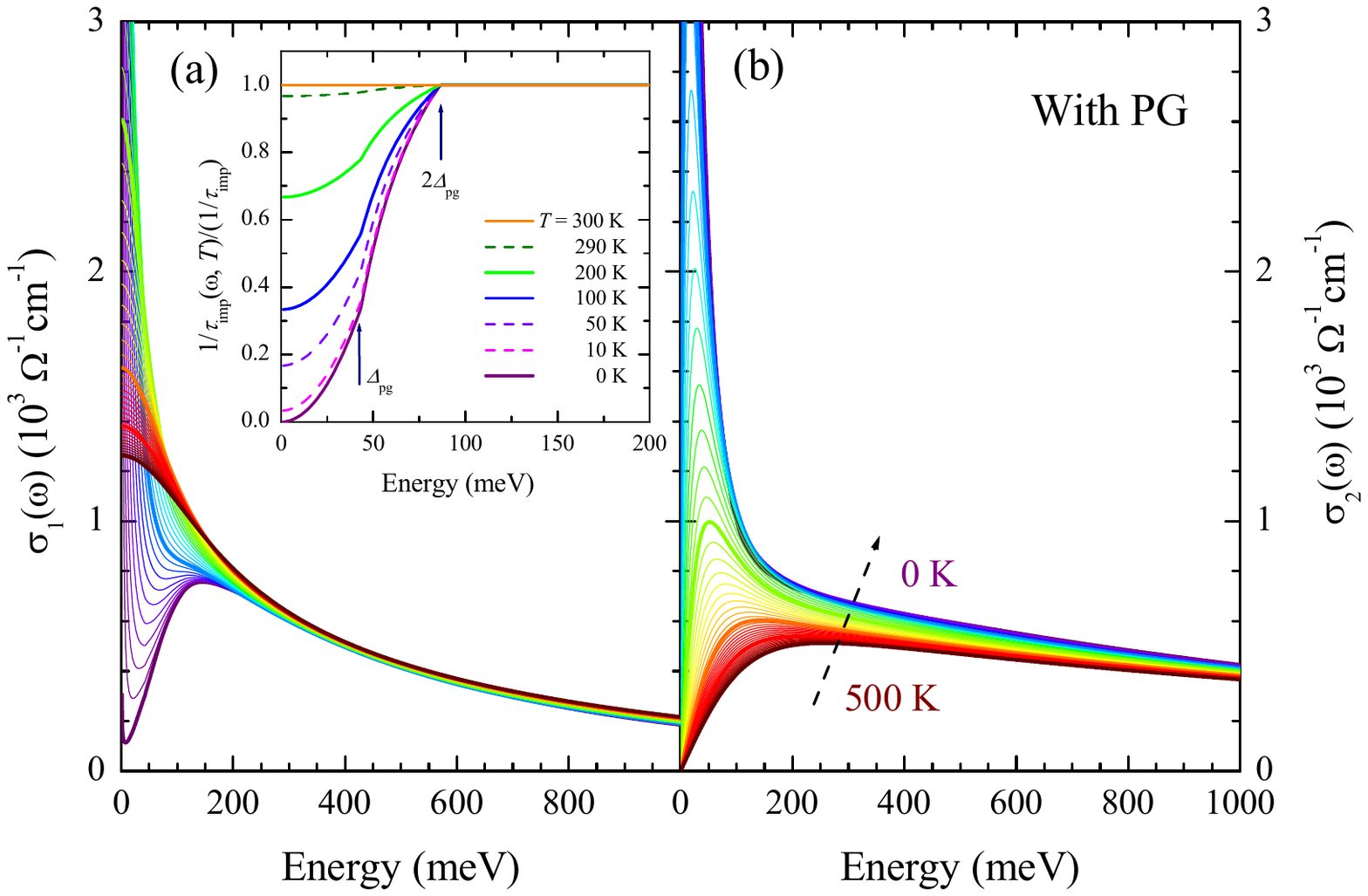}}%
  \vspace*{-0.5 cm}%
\caption{(Color online) Complex optical conductivity, including the pseudogap. (a) Real and (b) imaginary parts of the optical conductivity at temperatures ranging from 0 to 500 K with a 10 K increment. In the inset, the frequency- and temperature-dependent impurity scattering rates at several selected temperatures are shown. Note that the scattering rate is normalized by $1/\tau_{\mathrm{imp}}$. The optical conductivity is obtained from the optical self-energy using the extended Drude model with $\Omega_p$ of 2.0 eV and $1/\tau_{\mathrm{imp}}$ of 15 meV (see the text for a more detailed description).}
\label{fig3a}
\end{figure}

Figs. \ref{fig3a}(a) and (b) show the real and imaginary parts of the optical conductivity obtained from the real and imaginary parts of the optical self-energy using the extended Drude model. To get the optical conductivity, the plasma frequency ($\Omega_p$) was set at 2.0 eV, and the impurity scattering rate was included in $-2\Sigma_2^{op}(\omega)$. For the case of pseudogap, the impurity scattering rate is energy- and temperature-dependent \cite{hwang:2018} as shown in the inset of Fig. \ref{fig3a}(a), i.e., $1/\tau_{\mathrm{imp}}(\omega,T) = (1/\tau_{\mathrm{imp}})(1/\omega) \int_0^{\omega} N(x,T) dx$, where $N(x,T)$ is the pseudogap described in Eq. (3) and $1/\tau_{\mathrm{imp}} = $ 15 meV. Therefore, we included the scattering rate ($1/\tau_{\mathrm{imp}}(\omega,T)$) in $-2\Sigma_2^{op}(\omega,T)$ (see Fig. S2(a)in the Supplementary Material) and called the resulting optical scattering rate as a total scattering rate ($1/\tau^{op, \mathrm{Total}}(\omega) \equiv -2\Sigma^{op, \mathrm{Total}}_2(\omega)$). The total optical scattering rate was used to obtain the corresponding real part of the optical self-energy ($-2\Sigma_1^{op, \mathrm{Total}}(\omega,T))$ using the Kramers-Kronig relation (see Fig. S2(b) in the Supplementary Material). And then we obtained the complex optical conductivity ($\tilde{\sigma}(\omega, T)$) from the total optical self-energy ($-2\tilde{\Sigma}^{op, \mathrm{Total}}(\omega,T)$) using the extended Drude model. The real part of the optical conductivity shows a similar temperature-dependent trend as the measured $\sigma_1(\omega,T)$ in the published paper \cite{mirzaei:2013}. Here, it is worth noting that because at $T =$ 0 K, the total scattering rate is zero at $\omega =$ 0, i.e. $-2\Sigma^{op, \mathrm{Total}}_2(0) = -2\Sigma_2^{op}(0) + 1/\tau_{\mathrm{imp}}(0) =$ 0 (see Fig. S2(a) in the Supplementary Material), some amount of the spectral weight is confined at zero energy and appears as the Dirac $\delta$-function ($\delta(\omega)$) as shown in Fig. \ref{fig3a}(a). However, the fully gapped pseudogap at $T$ = 0 K cannot be a practical situation because the pseudogap is defined by a partial suppression in the density of states near the Fermi level. Even though $-2\Sigma^{op, \mathrm{Total}}_2(0) = 0$ at $T =$ 0, $\sigma_1(0)$ is not zero because $-2\Sigma^{op, \mathrm{Total}}_1(0)$ is also zero at $T =$ 0 (see Fig. S2(b) in the Supplementary Material). It should be noted that the pseudogap state is a normal state with an onset temperature ($T^*$) below which the pseudogap state appears. When the material enters the superconducting state below the SC transition temperature ($T_c$), the electrons are paired and condensed for the macroscopic SC ground state, and simultaneously the SC gap is opened. In this study, we only deal with the normal state, including the pseudogap state. Therefore, our discussions in this study are valid only at temperatures above $T_c$ in the case of material systems. Also, note that the total scattering rate ($-2\Sigma^{op, \mathrm{Total}}_2(\omega)$) shows similar $\xi^2$ ($\equiv (\hbar \omega)^2 +(p \pi k_B T)^2 $) dependencies as the optical scattering rate ($-2\Sigma^{op}_2(\omega)$) (see Fig. S4 in the Supplementary Material), where $\hbar$ is the reduced Planck constant, $k_B$ is the Boltzmann constant, and $p$ is an adjusting parameter, which may depend upon the material systems \cite{nagel:2012,maslov:2012,mirzaei:2013}. We obtained the amplitude and phase angle of the optical complex conductivity and showed them in Fig. \ref{fig2b}(c) and (d). The results seem to be consistent with those of underdoped cuprate in a published paper \cite{heumen:2022}.

\begin{figure}[!htbp]
  \vspace*{-1.1 cm}%
 \centerline{\includegraphics[width=7.0 in]{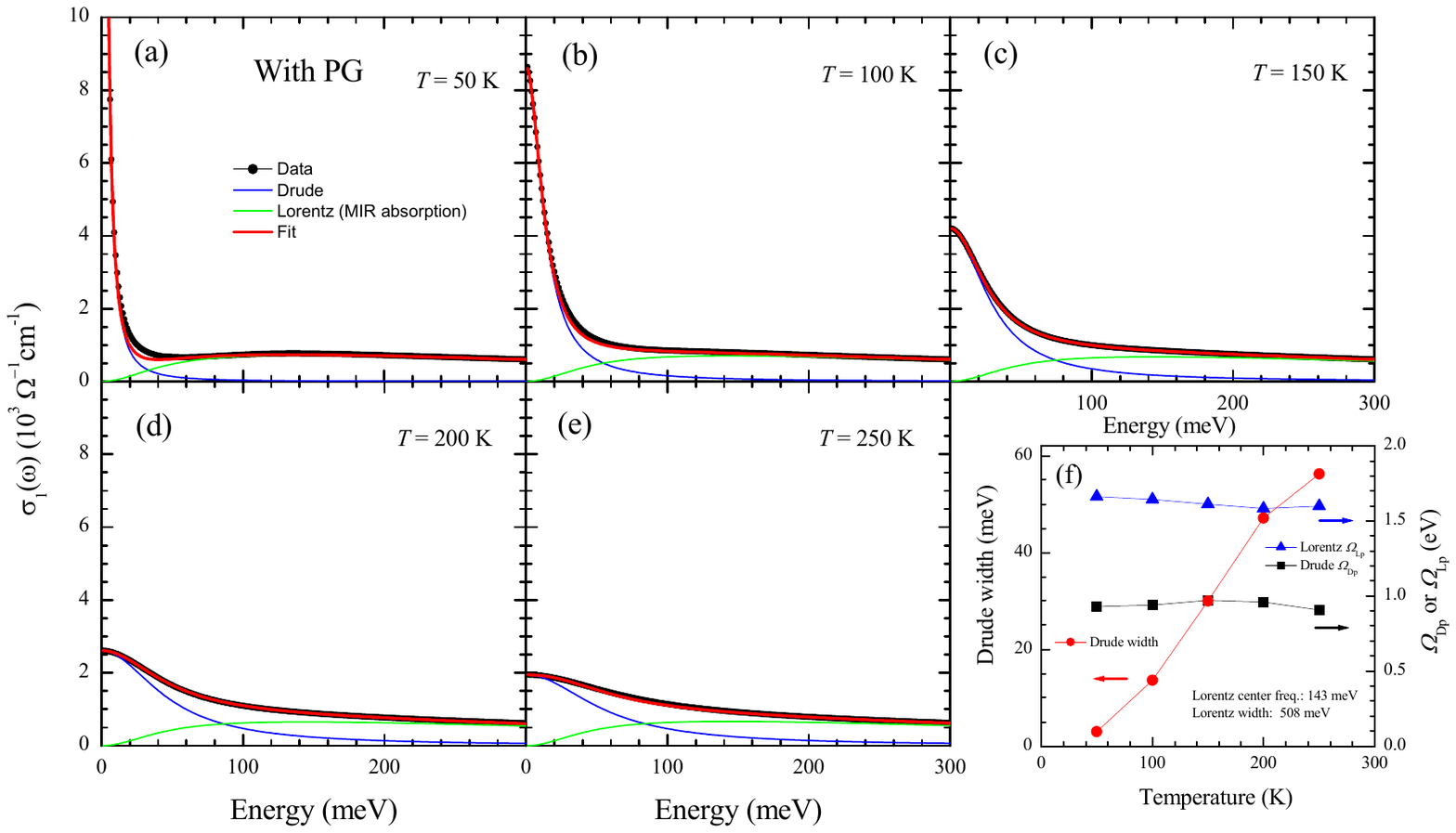}}%
  \vspace*{-1.7 cm}%
\caption{(Color online) (a-e) The optical conductivity data and Drude-Lorentz model fits at five selected temperatures in the PG phase. The data were fitted up to 300 meV with a Drude mode and a Lorentz mode. (f) The temperature-dependent fitting parameters are the Drude plasma frequency ($\Omega_{\mathrm{Dp}}$) and width, and the Lorentz center frequency, width, and plasma frequency ($\Omega_{\mathrm{Lp}}$). Note that the center frequency and width of the Lorentz mode are fixed. Both the Drude and Lorentz spectral weights show almost no temperature dependence.}
\label{fig3b}
\end{figure}

We fitted the optical conductivity data at five selected temperatures with the Drude-Lorentz model up to 300 meV and showed the results in Figs. \ref{fig3b}(a-e). As we can see in the figure, the conductivity consists of two components: a Drude component and a Lorentz one. The Lorentz one is known as the mid-infrared absorption (or the conformal tail) \cite{quijada:1999,heumen:2022,kumar:2022}. The results are consistent with those of underdoped cuprates \cite{heumen:2022,kumar:2022}; the quality of the fit is getting better as the temperature increases. In Fig. \ref{fig3b}(f), we show the fitting parameters as functions of temperature. The Lorentz mode is almost temperature-independent, whereas the Drude mode exhibits a strong $T$-dependency, particularly the width. These $T$-dependent trends of fitting parameters are consistent with those of underdoped cuprates in the literature \cite{kumar:2022}. We also fitted the optical conductivity at the same selected temperatures in the case without including the PG with the same scheme as in the reference [6], i.e., using an almost $T$-independent Lorentz mode and a $T$-dependent Drude model and showed the results in Fig. S3 in the Supplementary Material. The quality of the fits is not as good as in Fig. \ref{fig3b} and is getting worse as the temperature increases. Therefore, the results are not consistent with those of underdoped cuprates in the literature \cite{kumar:2022}.

\begin{figure}[!htbp]
  \vspace*{-0.1 cm}%
 \centerline{\includegraphics[width=7.0 in]{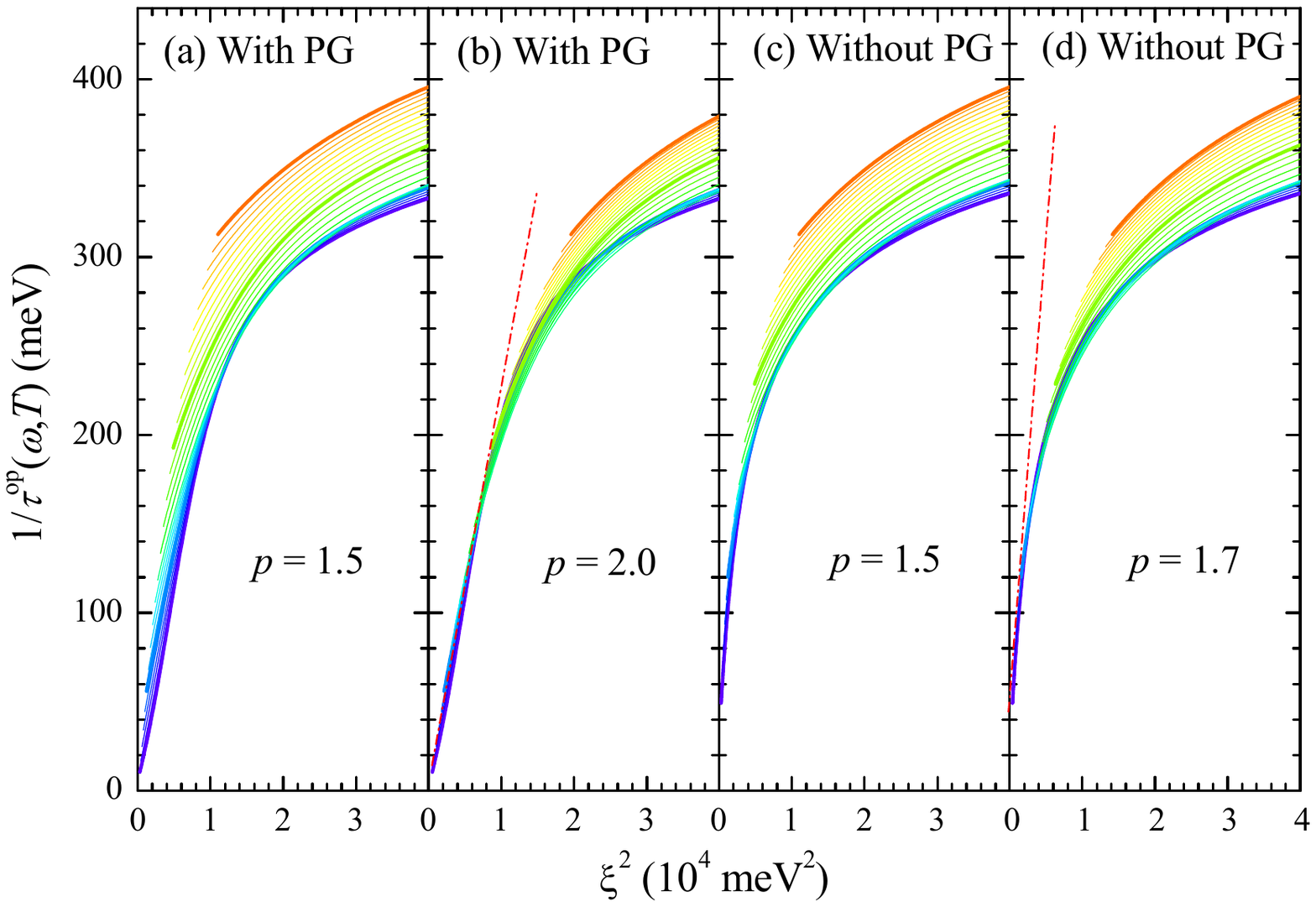}}%
  \vspace*{-0.3 cm}%
\caption{(Color online) Optical scattering rates as functions of $\xi^2 = (\hbar \omega)^2+(p \pi k_B T)^2$ for (a) $p =$ 1.5 and (b) $p =$ 2.0 including the PG and (c) $p =$ 1.5 and (d) $p =$ 1.7 without including the PG. Note that the thick blue curve is 100 K, the thick green curve is 200 K, and the thick orange curve is 300 K. The dash-dotted lines in (b) and (d) are guides for the eyes.}
\label{fig4}
\end{figure}

Furthermore, in Figs. \ref{fig4}(a) and \ref{fig4}(b), we plot the optical scattering rates as functions of $\xi^2$ in the case of including the pseudogap for $p =$ 1.5 and 2.0, respectively. Note that the optical scattering rates at temperatures between 50 and 300 K are shown for including only the pseudogap phase and reasonable pseudogap depths. When $p$ is 2.0, all curves of the optical scattering rates at various temperatures fall into a single curve at values of $\xi^2$ below $\sim$10000 meV$^2$ and are linear to $\xi^2$, indicating that the optical scattering rates show a Fermi liquid behaviour in the pseudogap phase. This result is very similar to the experimentally observed results of underdoped cuprate systems \cite{mirzaei:2013}. We could not find the reason why our data show a $p$-value of 2.0 rather than values less than 2 as in the published papers \cite{mirzaei:2013,kumar:2022} yet. It is worth noting that the optical scattering rate in this paper ($1/\tau^{op}(\omega) = -2\Sigma_2^{op}(\omega)$) is different from the optical scattering rate ($\hbar/\tau(\omega)$) in the literature by Kumar {\it et al.} \cite{kumar:2022}, i.e., $\hbar/\tau(\omega) = 1/\tau^{op}(\omega) (m_b/m^*(\omega))$, where $m_b$ is the band mass and $m^*(\omega)$ is the optical effective mass and $m^*(\omega)/m_b \equiv 1 + [-2\Sigma_1^{op}(\omega)]/\omega$. However, even though the optical scattering rates in the absence of the pseudogap fall into a single curve at low values of $\xi^2$ for $p =$ 1.5 and 1.7 (Figs. \ref{fig4}(c) and \ref{fig4}(d)), they are not linear to $\xi^2$. Therefore, they do not show a Fermi liquid-like behaviour.

We also performed the same calculations with a temperature-independent $I^2B(\Omega)$ (see Fig. S5 in the Supplementary Material). The $T$-independent coupling constant ($\lambda$) as a function of temperature is shown in the inset of Fig. S5. All the results are shown in Figs. S5-S12 in the Supplementary Material. The optical scattering rates (or imaginary parts of the optical self-energy) and the real parts of the optical self-energy (see Figs. S6(a) and (b) and S8(a) and (b)) differ significantly from the corresponding measured spectra \cite{hwang:2004,hwang:2006,hwang:2007a,mirzaei:2013}. The overall temperature-dependent exhibits a strong $T$-dependence and the $T$-dependent change of the imaginary part of the optical self-energy is significantly greater than that of the measured optical scattering rate \cite{mirzaei:2013}. We obtained the real part from the imaginary part using the Kramers-Kroing relation between them. We also showed the total optical self-energy in Fig. S9 for the case of including the PG, which are used to get the optical conductivity. We also obtained the optical conductivity spectra from the optical self-energy using the extended Drude model and showed them in Figs. S7 and S10. The real part of the optical conductivity in the case of including the PG (Fig. S10(a)) differs from the measured conductivity \cite{hwang:2006,hwang:2007a,mirzaei:2013}. We also showed the optical scattering rate as a function of $\xi^2$ in Fig. S11. The curves are very widely spread vertically, even though they fall into a single line at low $\xi^2$ values below $\sim$6000 meV$^2$ (Fig. S11(b)) for the case of including PG. We also show the total optical scattering rate as a function of $\xi^2$ in Fig. S12. These quantities differ from those observed experimentally \cite{mirzaei:2013}. These results support the fact that the $I^2B(\Omega)$ should be temperature dependent for simulating the experimentally observed temperate-dependent optical spectra using the procedure described in this paper.

Our results indicate that the observed Fermi liquid-like behaviour is intimately associated with both the $T$-dependent EBSD function and the $T$-dependent pseudogap. The two physical quantities directly contribute to the optical scattering rate. The EBSD function represents correlations in a correlated electron system such as cuprates and Fe-based superconductors, whereas the pseudogap reduces the density of states near the Fermi level, resulting in a partial gap near the Fermi level. Therefore, these two quantities give opposite contributions to the optical scattering rate. Generally, the EBSD function enhances the optical scattering due to the correlations (or scatterings) between electrons, whereas the pseudogap suppresses the optical scattering rate in the pseudogap region due to the reduction of scatterers. The EBSD function gives an up-step feature when one measures from zero energy to high energy, whereas the pseudogap gives a down-step feature when one measures from high energy above the pseudogap energy to low energy. Both give a step-like feature even though their line shapes are not the same. However, it is not easy to distinguish the two contributions clearly in a measured optical spectrum. In this paper, we separately show the two features in the optical scattering rate using the approach described in this paper. We also show that the two quantities are temperature dependent and the combined contributions of the two quantities give the Fermi liquid-like behaviour observed experimentally. Therefore, the observed Fermi liquid-like behaviour can be understood by considering the combined contributions from both the correlations between electrons and the pseudogap phenomenon.

There were some interesting optical studies on Fermi-liquid-like behavior and strange (or bad) metal state in the optical spectra of cuprates \cite{heumen:2022,kumar:2022}. These studies can be categorized into two groups based on the methods of analysis \cite{quijada:1999}: one is a one-component analysis, and the other is a two-component analysis. These analyses are also called the Mathiessen interpretation (one-component analysis) and anti-Mathiessen interpretation (two-component analysis) \cite{heumen:2022}. In these previous works \cite{heumen:2022,kumar:2022}, the authors used the two analyses to analyze the measured optical spectra in a wide doping range, from underdoped to overdoped, whereas in our study, we simulated the optical self-energy and optical conductivity of an underdoped cuprate starting from the $T$-dependent EBSD function using the generalized Allen's formula and extended Drude model. The optical scattering rate (the imaginary part of the optical self-energy) exhibits Fermi-liquid-like behavior (see Fig. \ref{fig4}), and the optical conductivity shows two components: the Drude component and the mid-infrared absorption (or conformal tail) (see Fig. \ref{fig3b}). The $T$-dependent evolutions of these two components are consistent with those of measured underdoped cuprates \cite{heumen:2022,kumar:2022}. Additionally, in our approach, the correlations between electrons were explicitly revealed through the EBSD function.

\section{Conclusion}

We investigated the Fermi liquid-like behaviour observed by optical spectroscopy\cite{mirzaei:2013} in the pseudogap phase of underdoped cuprates. We applied the reverse process of a well-established approach, which has been used to extract the EBSD function from the measured optical spectrum, to obtain the optical scattering rate in the pseudogap phase of cuprate systems. We successfully simulated the observed Fermi liquid-like behaviour using the reverse process with the temperature-dependent EBSD function and Fermi arc modelled pseudogap. Therefore, the observed Fermi liquid-like behaviour can be understood by considering the combined contributions from the $T$-dependent EBSD function and the $T$-dependent pseudogap. Additionally, we obtained optical conductivity from the optical self-energy using the extended Drude model. The optical conductivity in the low-energy region shows two components: a Drude component and a Lorentz one. The Lorentz component is almost $T$-independent, whereas the Drude one is strongly $T$-dependent, particularly the width. These results are consistent with previous optical studies \cite{heumen:2022,kumar:2022}. Furthermore, our results indicate that the $T$-dependent EBSD functions extracted from the measured optical spectra of cuprates using the established approach are intrinsic and contain information regarding the glue spectrum for the Cooper pair formation.
\\

\noindent {\bf Acknowledgements} This paper was supported by the National Research Foundation of Korea (NRFK Grant Nos. 2020R1A4A4078780 and 2021R1A2C101109811).

%
%
\bibliographystyle{naturemag}
\bibliography{bib}

\end{document}